 \documentstyle{article}

\newfont{\g}{eufm9}

\newcommand{\gtg}{\mbox{\g g}}

\newcommand{\hgtg}{\mbox{$\hat{\gtg}$}}

\newcommand{\gtsl}{\mbox{\g sl}}

\newcommand{\gth}{\mbox{\g h}}
\newcommand{\nc}{\mbox{${\bf C}$}}

\newcommand{\cP} {\mbox{${\bf CP}$}}

\newcommand{\cp}{\mbox{${\cal P}$}}

\newtheorem{theorem}{Theorem}[section]

\newtheorem{lemma}[theorem]{Lemma}
\begin{titlepage}
\title{ }
\end{titlepage}

\title{{ \bf Solutions to Knizhnik-Zamolodchikov equations with coefficients
in non-bounded modules}}
\author{Kenji Iohara\thanks{e-mail address: iohara@kurims.kyoto-u.ac.jp}
, Feodor Malikov\thanks{Supported by the
Japan Society for the Promotion of Science Post -Doctoral Fellowship for
Foreign Researchers in
Japan.} \thanks{Address after August, 1, 1993: Department of Mathematics Yale
University New Haven
CT 06520 USA; e-mail address: malikov@kusm.kyoto-u.ac.jp}\\
 Department of Mathematics, Kyoto University ,\\
 Kyoto 606 Japan }
\date{Received: }
 
\begin{document}

\maketitle

\begin{abstract}
We explicitly write dowm integral formulas for solutions to
Knizhnik-Zamolodchikov equations
with coefficients in non-bounded -- neither highest nor lowest weight --
$\gtsl_{n+1}$-modules.
The formulas are closely related to WZNW model at a rational level.
\vspace{5 mm}

\end{abstract}
\section {{\bf Introduction}}
\pagestyle{myheadings}
\markright{ Solutions to KZ equations}

 Let $\gtg$ be a finite-dimensional simple Lie algebra, $\hgtg$ the
corresponding non-twisted affine
Lie algebra. Let $\lambda$ be a weight of $\gtg$ ,
$M(\lambda,\,  k)$ ($M(\lambda,\, k)^{c}$ ) be  the Verma ( contragredient
Verma) module
over $\hgtg$ with the central charge $k$;
for a $\gtg-$module $V$ be  denote by $V((z))$ the module of formal Laurent
series in $z$ with coefficients
in $V$, regarded as a $\hgtg-$module with the central charge equal to 0.

Vertex operator is a $\hgtg-$linear map
\begin{equation}
\label{def_of_vert_op}
\Phi (z):\; M
(\lambda_{1},\,  k)\rightarrow
 M(\lambda_{2},\, k)^{c}\otimes V((z)).\end{equation}

If highest weights $(\lambda_{1},k),\cdots(\lambda_{N+1},k)$ are generic then
\newline
\mbox{$M(\lambda_{i},k)\approx M(\lambda_{i},k)^{c},\;1\leq i\leq N+1$} and one
may consider a product
of vertex operators $\Phi_{N}(z_{N})\circ\cdots\circ\Phi_{1}(z_{1})$.  Matrix
element
 $\langle v_{\lambda_{N+1}}^{\ast},\,
\Phi_{N}(z_{N})\circ\cdots\circ\Phi_{1}(z_{1})
v_{\lambda_{1}}\rangle$ related to vacuum vectors is
called a correlation function. One of the central results of conformal field
theory  ( see ~\cite{kn_zam}) is that a correlation function satisifies a
remarkable system of
Knizhnik-Zamolodchikov equations. We prepare notations in order to write down
the trigonometric form
of Knizhnik-Zamolodchikov equations.

 Let
\[\gtg=\gth\oplus\oplus_{\alpha\in\Delta}\gtg_{\alpha}\] be a root space
decomposition.
 Fix an invariant inner product on $\gtg$ and a basis
$\{h_{i}\in\gth,\;g_{\alpha}\in\gtg_{\alpha}:\;1\leq i\leq
n,\;\alpha\in\Delta\}$ of $\gtg$  so that
$(h_{i},h_{j})=\delta_{i,j},\;(g_{\alpha},g_{\beta})=\delta_{\alpha,-\beta}$.
For each
$\mu\in\gth^{\ast}$ denote by $h_{\mu}$ an element of $\gth$ satisfying (and
uniquely determined) by
the condition $(h_{\mu},h)=\mu (h)$.

 Set
\[r=\frac{1}{2}\sum_{i=1}^{n}h_{i}\otimes
h_{i}+\sum_{\alpha\in\Delta_{+}}g_{\alpha}\otimes
g_{-\alpha}.\]
Being an element of $U(\gtg)\otimes U(\gtg)$ $r$ naturally acts on a tensor
product of
 2 $\gtg-$modules. There are $N^{2}$ different ways to make it act on a tensor
product of $N$
$\gtg-$modules via the following $N^{2}$ embeddings of $U(\gtg)^{\otimes 2}$ in
 $U(\gtg)^{\otimes N}$: each of them is associated to a pair of numbers $1\leq
i,j\leq N$ and sends
\[U(\gtg)^{\otimes 2}\ni \omega\mapsto \omega_{ij}\in U(\gtg)^{\otimes N},\]
so that
\[\mbox{if } \omega =
\sum_{s}a_{s}\otimes b_{s}\mbox{ then }\omega_{ij}=\sum_{s}
\overbrace{\underbrace{1\otimes\cdots 1\otimes a_{s}}_{i}\otimes 1\otimes\cdots
\otimes 1\otimes
b_{s}}^{j}\otimes 1\otimes\cdots\otimes 1.\]

For a pair $1\leq i,j\leq N$
introduce the following function in 2 complex variables with values in
$U(\gtg)^{\otimes N}$:
\[r(z_{i},z_{j})=\frac{r_{ij}z_{i}+r_{ji}z_{j}}{z_{i}-z_{j}}.\]

\begin{theorem}[Knizhnik, Zamolodchikov]
The correlation function
\[\Psi (z)= \langle v_{\lambda_{N+1}}^{\ast}\,
\Phi_{N}(z_{N})\circ\cdots\circ\Phi_{1}(z_{1})
v_{\lambda_{1}}\rangle\]
satisfies the following system of
differential equations

\begin{eqnarray}
\label{k_z_class}
(k+h^{\vee} )
z_{i}\frac{\partial\Psi}{\partial z_{i}}&=&
\{\sum_{j\neq
i}r_{ij}(z_{i},z_{j})-\frac{1}{2}(\lambda_{1}+\lambda_{N+1}+2\rho)^{(i)}\}\Psi,\\
1\leq &i&\leq N,\nonumber
\end{eqnarray}
where $
h^{\vee}$ is the dual Coxeter number of $\gtg$ and for each $\mu
\in\gth^{\ast}$ $\mu^{(i)}$ stands for the  operator acting on
$V_{1}\otimes\cdots\otimes V_{N}$ as $h_{\mu}$ applied to the $i-$th factor of
$V_{1}\otimes\cdots\otimes V_{N}$.
\end{theorem}
 To keep track of the parameters we will be referring to
(\ref{k_z_class}) as $KZ(\lambda_{N+1},\lambda_{1})$.

 The deep theory of KZ equations has been developed by several authors (see
e.g.
{}~\cite{ts_kan,sch_1,sch_varch_1,sch_varch_2} ) in the case when $V_{i}$ are
highest weight modules.
It has also been realized that this theory is relevant to physics applications
in the case when
$(\lambda_{i},k)$ is either integral or generic.Indeed, if conflicting
with the above assumptions,  some of $M(\lambda_{i},k)$ are reducible then the
product
$\Phi_{N}(z_{N})\circ\cdots\circ\Phi_{1}(z_{1})$ does not exist unless each of
the operators
$\Phi_{i}(z_{i})$ can be pushed down to a map
\begin{equation}
\label{pushed_down_vert_op}
\Phi_{i} (z_{i}):\; L
(\lambda_{i},\,  k)\rightarrow
 L(\lambda_{i+1},\, k)\otimes V_{i}((z_{i})),
\end{equation}
where $L(\lambda,k)$ stands for an irreducible highest weight module with the
highest weight
$(\lambda,k)$. In the case when $(\lambda,k)$ is an admissible weight (for
example, dominant integral
weight)  ~\cite{kac_wak} the
last condition reduces to the {\em singular vector decoupling condition}:
matrix elements of
$\Phi_{i}(z_{i})$ related to singular vectors of $M(\lambda_{i})$ vanish.
It is known ~\cite{feig_sch_varch}
 that if
each $(\lambda_{i},k)$ is dominant integral
 then everything goes through nicely, in particular, the Schechtman-Varchenko
integral solutions to (\ref{k_z_class}) come from  products of vertex operators
(\ref{pushed_down_vert_op}). However if the
central charge
is not integral it has been realized (~\cite{awata}, see also
{}~\cite{f_g_p_p_1,f_g_p_p_2})
that the singular vector decoupling condition
implies that $V_{i}$ is neither highest nor lowest weight module. Though some
results for such
models were obtained in ~\cite{f_g_p_p_1,f_g_p_p_2}, where in particular the
connection to quantum
hamiltonian reduction was revealed, not much is known about KZ equations in
this case.

In ~\cite{feig_mal} a new method of constructing solutions to (\ref{k_z_class})
was proposed
which seems to be relevant to the problem. Let $G$ be a
complex Lie group related to $\gtg$, $F=G/B$ be a flag manifold and
$F^{0}\subset F$ be the big cell.
There is a family of embeddings of $\gtg$ into the algebra of order 1
differential operators
 on $F^{0}$
\[\pi_{\mu}:\gtg\rightarrow Diff^{1}(F^{0}),\;\mu\in\gth^{\ast}.\]
 This makes the space of analytic
 functions on $F^{0}$ into a huge $\gtg-$module. Different $\gtg-$closed
subspaces give realization
of different $\gtg-$modules. For example, contragredient Verma modules are
realized in the space of
polynomials on $F^{0}$, $\mu$ being the highest weight and a constant function
being a highest
weight vector; this observation has been extensively used recently with regards
to Wakimoto
modules ~\cite{feig_fren_0,bowkn,aw_ts_yam}.
  The spaces of multi-valued functions give modules with quite different
properties, the
simplest example being that of $\gtsl_{2}$: in this case the big cell is $\nc$,
 contragredient
Verma modules are realized in $\nc[x]$; the space
$x^{\nu}\nc[x,x^{-1}],\;\nu\in\nc$ is also closed
under the action
of $\gtsl_{2}$ and the embedding $\pi_{\mu},\;\mu\in\nc$ makes it into
generically irreducible
$\gtsl_{2}-$module. This module is transparently neither highest nor lowest
weight one.

Regarding $V$ in (\ref{def_of_vert_op}) as a $\gtg-$ module realized in
functions on the big cell
one identifies elements of $V((z))$ with functions of 2 groups of variables:
$x$ and $z$, where
$x$ stands for a ( vector ) coordinate on the big cell and $z$ is a coordinate
on $\nc$. Likewise,
the correlation function
\[\Psi (z)= \langle v_{\lambda_{N+1}}^{\ast},\,
\Phi_{N}(z_{N})\circ\cdots\circ\Phi_{1}(z_{1})
v_{\lambda_{1}}\rangle\]
is identified with a function of $x^{(1)},\ldots,x^{(N)};z_{1},\ldots,z_{N}$
where $x^{(i)}$ is a
coordinate on the $i-$th copy of the big cell, $z_{i}\in\nc,\;1\leq i\leq N$.
One of the advantages of this functional realization is that the emebedding
$\pi_{\lambda}:\gtg\rightarrow Diff^{1}(F^{0})$ lifts to the mapping of the
group $G$: for
$g\in\gtg$ the exponent $\exp (-tg)$ is a well-defined operator.

Let \(W\) be the Weyl group of \(\widehat{\gtg}\),
\(w=r_{m_{1}}r_{m_{2}}\cdots r_{m_{l}}\in W\) be a  decomposition(not
necessarily
reduced),where \(r_{m}\) denotes the  reflection at the corresponding simple
root. Set,
\[\beta_{j}=\frac{2(r_{m_{l+2-j}}\cdots r_{m_{l}}\cdot
\lambda_{1},\alpha_{m_{l+1-j}})}
                  {(\alpha_{m_{l+1-j}},\alpha_{m_{l+1-j}})}+1,\;1\leq j \leq l
.\]

Given
\[\Psi_{old} (z)= \langle v_{\lambda_{N+1}}^{\ast},\,
\Phi_{N}(z_{N})\circ\cdots\circ\Phi_{1}(z_{1})
v_{\lambda_{1}}\rangle,\]

set
\begin{equation}
\label{our_solution_gener}
\Psi_{new}=\prod_{j=1}^{l}\Gamma(-\beta_{j})^{-1} \int
\{\exp(-t_{l}F_{m_{l}}) \cdots
\exp(-t_{1}F_{m_{1}})\Psi_{old}\}\prod_{j=1}^{l}t_{j}^{-\beta_{j}-1}\,
dt_{1}\cdots dt_{l},
\end{equation}
where the integration is carried out over an arbitrary cycle of the highest
homology group
related to the multi-valued integrand. In (\ref{our_solution_gener}) it is set
that
$E_{i},F_{i},H_{i},\;0\leq i\leq rk\,\gtg$ are canonical Cartan generators of
$\hgtg$ and
$E_{i},F_{i},H_{i},\;1\leq i\leq rk\,\gtg$ are the ones coming from the
inclusion
$\gtg\subset\hgtg$.

\begin{theorem} \mbox{~\cite{feig_mal}} $\Psi_{new}$ is a solution to
$KZ(\lambda_{N+1},w\cdot\lambda_{1})$. \label{our_result}
\end{theorem}

Theorem~\ref{our_result} works as follows:  given a solution to KZ it generates
new ones labelled by
elements of the affine Weyl group. In our notations the simplest solution to
$KZ(\lambda_{N+1},\lambda_{1})$ is given by
\begin{eqnarray}
\stackrel{\circ}{\Psi}_{old} =&\prod_{i<j}&(z_{i}-z_{j})^{2(\mu_{i},\mu_{j}) /
(k+h^{\vee})}(z_{i}z_{j}) ^{-(\mu_{i},\mu_{j}) / (k+h^{\vee})}\times\nonumber\\
&\prod_{i}&z_{i}^{(\lambda_{1}+\lambda_{N+1}+2\rho , \mu_{i}) /
2(k+h^{\vee})},\nonumber
\end{eqnarray}
where $\mu_{i}$ is a highest weight of $V_{i}$, $1\leq i\leq N$.
In particular, $\stackrel{\circ}{\Psi}_{old}$ is independent of $x'$s.
The purpose of this paper
 is to explicitly write down the integral
\begin{eqnarray}
& &\Psi_{new}=\stackrel{\circ}{\Psi}_{old}\times\nonumber\\
& &\prod_{j=1}^{l}\Gamma(-\beta_{j})^{-1} \int
\{\exp(-t_{l}F_{m_{l}}) \cdots
\exp(-t_{1}F_{m_{1}})1\}\prod_{j=1}^{l}t_{j}^{-\beta_{j}-1}\,
dt_{1}\cdots dt_{l},
\label{what_to_calc}
\end{eqnarray}
for $\gtg=\gtsl_{n+1}$, generalizing the calculation carried out in
{}~\cite{feig_mal} for $\gtsl_{2}$.

{\bf Remark.} The integral representation of $\Psi_{new}$ in
Theorem~\ref{our_result} is nothing but
the conventional definition of $F_{m_{l}}^{\beta_{l}}\cdots
F_{m_{1}}^{\beta_{1}}\cdot\Psi_{old}$.
The latter comes from looking at the ``matrix element''
\[\langle v_{\lambda_{N+1}}^{\ast}\,
\Phi_{N}(z_{N})\circ\cdots\circ\Phi_{1}(z_{1})
F_{m_{1}}^{\beta_{1}}\cdots F_{m_{l}}^{\beta_{l}}
v_{\lambda_{1}}\rangle.\]
Though the expression $F_{m_{1}}^{\beta_{1}}\cdots F_{m_{l}}^{\beta_{l}}
v_{\lambda_{1}}$ is not understood as an element of $M(\lambda_{1},k)$,
the powers are chosen in such a way that it formally satisfies the
singular vector conditions ~\cite{feig_mal,malff}, which makes the statement of
Theorem~\ref{our_result} almost obvious. One can similarly consider an
expression
\[\langle E_{m_{1}}^{\beta'_{1}}\cdots E_{m_{l}}^{\beta'_{l}}\cdot
 v_{\lambda_{N+1}}^{\ast}\, \Phi_{N}(z_{N})\circ\cdots\circ\Phi_{1}(z_{1})
v_{\lambda_{1}}\rangle,\]
for appropriate $\beta'_{1},\ldots\beta'_{l}$ and write down another solution
in the form
close to (\ref{our_solution_gener}) but with $F'$s replaced with $E'$s or
combine both methods
or, finally, apply them to other solutions obtained in
{}~\cite{sch_varch_1,f_g_p_p_1,f_g_p_p_2}.

As to relation of our solution (\ref{our_solution_gener})
to correlation
functions, we have been able to verify in simplest cases that
(\ref{our_solution_gener}) indeed gives
a matrix element of a product of vertex operators and hope that
(\ref{our_solution_gener})
 will prove useful
for investigation of other rational level models.

{\bf Acknowledgements.} Our thanks are due to M.Jimbo for his interest in the
work. Results of the
work were announced when F.M. visited the National Laboratory for High Energy
Physics in Tsukuba.
F.M. is obliged to H.Awata and Y.Yamada for their hearty hospitality
and interesting discussions.

\section{{\bf Integral formulas for soilutions of $KZ(\lambda_{N+1},
\lambda_{1})$   }}

\subsection{{\bf Main result}}
Here we are going to write down the integral (\ref{what_to_calc}) in the case
of $\gtg=\gtsl_{n+1},\;
\hgtg =\widehat{\gtsl_{n+1}}$. In this case there are $3(n+1)$ Cartan
generators
$E_{i},F_{i},H_{i},\;0\leq i\leq n$, where
$E_{i},F_{i},H_{i},\;1\leq i\leq n$ are the ones coming from the inclusion
$\gtg\subset\hgtg$. Explicitly the generators are described as follows. If
$e_{ij}=(a_{st})$ is
an $(n+1)\times (n+1)$ matrix then $E_{i}=e_{ii+1},\;F_{i}=e_{i+1i},\;
H_{i}=e_{ii}-e_{i+1i+1}
,\; 1\leq i\leq n$ and
$E_{0}=e_{n+11}\otimes z,\;F_{0}=e_{1n+1}\otimes z^{-1}$ ( see ~\cite{kac_book}
for details).
 The $\gtg-$weight $\mu$ is considered as a vector $(\mu_{1},\ldots,
\mu_{n}),\;\mu_{i}=\mu(H_{i})$.
The embedding
\[\pi_{\mu}:\; \gtsl_{n+1}\rightarrow Diff^{1}(F^{0}),\;\mu=(\mu_{1},\ldots,
\mu_{n})\]
is calculated in ~\cite{feig_fren_0} (see also ~\cite{bowkn}).
To recall this result we choose  coordinates of the big cell $F^{0}$ to be
 $\{x_{ij}:\,1\leq i<j\leq n\}$ identifying as usual the big cell with the
subgroup of
matrices
\[\left( \begin{array}{lllll}
             1        & x_{11}  & \cdots & \cdots & x_{1 n} \\
                      & 1       & \ddots &        & \vdots  \\
                      &         & \ddots & \ddots & \vdots  \\
                      &         &        & \ddots & x_{n n} \\
             0        &         &        &        & 1       \\
         \end{array} \right). \]

 For $1 \leq i \leq n$ set
\[ \partial_{x_{i j}}:=\frac{\partial}{\partial x_{ij}}.\]

Then $\pi_{\mu}$ acts on Cartan generators by
\[  E_{i}\mapsto -\partial_{x_{i i}}-
           \sum_{j=i+1}^{n}x_{i+1 j}\partial_{ x_{i j}} , \]
  \[ F_{i}\mapsto x_{i i}\left(\sum_{j=1}^{i}x_{j i}\partial_{x_{j
i}}-\sum_{j=1}^{i-1}
          x_{j i-1}\partial_{x_{j i-1}}\right) -\sum_{j=i+1}^{n}x_{i
j}\partial_{x_{i+1,j}}
          +\sum_{j=1}^{i-1}x_{j i}\partial_{x_{j i-1}}+\mu_{i}x_{i i} . \]
Here $x_{i j}=0$ unless $1 \leq i \leq j \leq n$.

The matrix  $e_{1n+1}$ may be written as
\[ e_{1n+1}:=[ \cdots [E_{1},E_{2}],
      \cdots ],E_{n}] . \]
Using the above formulas one proves the following

\begin{lemma}
\[\pi_{\mu}( e_{1n+1})=-\partial_{x_{1 n}} \]
\end{lemma}
{}From now on till the end of this section we omit writing $\pi_{\mu}$
identifying Lie algebra
elements with their images under $\pi_{\mu}$.

The action of the Lie algebra $\hgtg$ on a function on $ F^{0}\times
\nc^{\ast}$
is determined by the evaluation map $g\otimes z^{k}\mapsto z^{k}g$. In
particular

\[ F_{0}=e_{1n+1}\otimes z^{-1}\mapsto -z^{-1}\partial_{x_{1 n}} .\]

The result of exponentiation of these formulas is given by

\begin{lemma}.

\label{g-action}

(i) If $\mu = 0$ then
 \[  \begin{array}{l}
   1) \exp (-tF_{0}):x_{k l} \longmapsto
    \left\{ \begin{array}{ll}
             z^{-1}t+x_{1 n} & \mbox{(k,l)=(1,n)} \\
             x_{k l}         & \mbox{otherwise}   \\
            \end{array} \right.\\
   2) \exp (-tF_{i}):x_{k l} \longmapsto
    \left\{ \begin{array}{ll}
             \frac{x_{k i}}{1+x_{i i}t} & \mbox{for $l=i$} \\
             -\left( x_{k i}-x_{k i-1}x_{i i} \right) t +x_{k i-1} & \mbox{for
$l=i-1$} \\
             x_{i l}t+x_{i+1 l} & \mbox{for $k=i+1$} \\
             x_{k l} & \mbox{otherwise} \end{array} \right. \\
  \end{array} \]

(ii) Generically
\[ \exp(-tF_{i}).\psi(x)=(1+x_{i i}t)^{\mu_{i}}\psi(x'),\;1\leq i\leq n \]
where $x'$ is given by the substitution of the item (i) while action of $F_{0}$
is independent of
$\mu$.

\end{lemma}

{\bf Proof}

If $\mu =0$ then all $F$'s are vector fields. The problem of evaluating an
exponent of a vector
field is, actually, a problem of the theory of ordinary differential equations:
the exponent of
a vector field is an element of a 1-parametric family of diffeomorphisms
generated by the vector
field and, therefore, is given by a general solution to the corresponding
system of o.d.e.'s.
In our case the system and the solution are (resp.):

1):
\[\dot{x}_{k l} =\delta_{k1}\delta_{ln}z^{-1}  \Longrightarrow  x_{1n}=z^{-1}t
+x_{1n}; \]
 2):
 \[ \begin{array}{lll}
     \dot{x}_{i i} = x_{i i}^2 & \Longrightarrow & x_{i i}=\frac{x_{i
i}(0)}{1-x_{i i}(0)t} \\
    \dot{x}_{j i} = x_{i i}x_{j i} & \Longrightarrow & x_{j i}=\frac{x_{j
i}(0)}{1-x_{i i}(0)t} \\
    \dot{x}_{j i-1} = x_{j i}-x_{j i-1}x_{i i} & \Longrightarrow
    & x_{j i-1}=\left( x_{j i}(0)-x_{j i-1}(0)x_{i i}(0) \right) t+x_{j i-1}(0)
\\
    \dot{x}_{i+1 j}=-x_{i j} \; \dot{x_{i j}}=0 \; \mbox{for $j \geq i+1$}
    & \Longrightarrow & x_{i j}=x_{i j}(0) \\
     & \Longrightarrow & x_{i+1 j}=-x_{i j}(0)t+x_{i+1 j}(0) \\
   \end{array}, \]
which completes proof of the item (i). As to the item (ii), one shows that any
order 1 diffrential
operator is conjugated to a vector field by the multiplication by a function it
annihilates. This
implies (ii) since $ F_{i}\cdot ( x_{ii}^{-\mu_{i}})=0$.
 {\bf Q.E.D.}

Now by using all this one can calculate the integrand of
(\ref{what_to_calc}).But to formulate the
result it is  convenient to give some more notations.

Set
\[ T=(t_{i j}):= \left(\begin{array}{llll}
                   x_{1 1} & \cdots & \cdots & x_{1 n} \\
                      1    & \ddots &        & \vdots  \\
                           & \ddots & \ddots & \vdots  \\
                      0    &        &    1   & x_{n n} \\
                  \end{array} \right) ;\]
 for $1\leq i_{k} \leq i_{k-1} \leq \cdots \leq i_{1} \leq j$ , $j+k\leq n+1$,
\[ I_{i_{1},i_{2},\cdots ,i_{k}}^{j}:=\{j+1-i_{1},j+2-i_{2},\cdots ,j+k-i_{k}\}
\]
\[  J_{k}^{j}:=\{j,j+1,\cdots ,j+k-1\} \]
\[ T_{i_{1},i_{2},\cdots i_{k}}^{j}:=(t_{i j})_{i \in I_{i_{1},i_{2},\cdots
,i_{k}}^{j}
                                               ,j \in J_{k}^{j}} \]
\[ Q_{i_{1},i_{2},\cdots i_{k}}^{j}:=\left\{\begin{array}{cl}
                           \det(T_{i_{1},i_{2},\cdots i_{k}}^{j}) & \mbox{for
$k>0$} \\
                                          1                       & \mbox{for
$k=0$} \\
                           \end{array} \right. \]

Introduce a collection of functions on the big cell along with an ordering on
it.

{\bf Definition} We write
\[ \begin{array}[t]{ll}
 & \begin{array}[t]{l}
     Q_{i_{1},i_{2}, \cdots ,i_{k}}^{j}  \stackrel{F_{l}}{\longrightarrow}Q'\\
    \stackrel{\rm def}{\Longleftrightarrow}
    \exp(-tF_{l})Q_{i_{1},i_{2},\cdots ,i_{k}}^{j}=\left\{\begin{array}{cl}
 \frac{1}{1+x_{j j}t}\left\{ Q't+Q_{i_{1},i_{2},\cdots ,i_{k}}^{j} \right\} &
\mbox{for $l=j$} \\
                             Q't+Q_{i_{1},i_{2},\cdots ,i_{k}}^{j}
&\mbox{for $l\neq j$}\\
                     \end{array} \right. \\
     \end{array}  \\

   \end{array} \]
In the definition it is assumed  that $\mu =0$.

\begin{lemma}
\label{action}
\[ \begin{array}{llllc}

1) & Q_{i_{1},i_{2},\cdots ,i_{k}}^{j} &
\stackrel{F_{j+r-1-i_{r}}}{\longrightarrow}
   & Q_{i_{1},i_{2},\cdots ,i_{r}+1,\cdots ,i_{k}}^{j} & \mbox{for $1\leq r
\leq k$} \\
2) & Q_{i_{1},i_{2},\cdots ,i_{k}}^{j} & \stackrel{F_{j+k}}{\longrightarrow}
   & Q_{i_{1},i_{2},\cdots ,i_{k},1}^{j} &\mbox{for $k\geq 0$} \\
3) & Q_{j,i_{2},\cdots ,i_{n+1-j}}^{j} & \stackrel{F_{0}}{\longrightarrow}
   & (-1)^{n-j}Q_{i_{2}-1,\cdots ,i_{k'}-1}^{j}z^{-1}
   & \mbox{where}  \\
   & & & &k'-1:=\sharp \{r:\;r>2,i_{r}>1\}\\
   \end{array} \]
\[ \mbox{Otherwise,$Q \stackrel{F_{l}}{\longrightarrow}0$} \] \end{lemma}

The proof of this lemma is a standard calculation of linear algebra using Lemma
\ref{g-action};
in particular we use Laplace expansion of a certain determinant to prove 2).

The above definition suggests to  introduce the following
$n+1-$colored graph $\Gamma$. The set of vertices of $\Gamma$ is the set of all
 $Q\neq 0$ such that
\[1 \stackrel{F_{j_{1}}}{\longrightarrow} Q_{1}
\stackrel{F_{j_{2}}}{\longrightarrow} Q_{2}
      \longrightarrow \cdots \longrightarrow Q_{r-1}
\stackrel{F_{j_{r}}}{\longrightarrow}, Q\]
for some $j_{1},\ldots,j_{r}$. It follows from Lemma~\ref{action} that each
vertex is of
the form $(-1)^{(n-j)r}Q_{i_{1},i_{2},\cdots i_{k}}^{j}z^{-r}$. Define a
function on the set of
vertices by
\[ l((-1)^{(n-j)r}Q_{i_{1},i_{2},\cdots
i_{k}}^{j}z^{-r})=(n+1)r+\sum_{p=1}^{k}i_{p}. \]

 Two vertices $P,Q$ are connected by an edge of the
color $i$ if and only if
\[P \stackrel{F_{i}}{\longrightarrow} Q.\]
With any vertex $Q\in \Gamma$ associate a set $\cp (Q)$ of all oriented paths
connecting 1 and $Q$.

\begin{lemma}
\label{length}
All $\gamma\in\cp ((-1)^{(n-j)r}Q_{i_{1},i_{2},\cdots i_{k}}^{j}z^{-r})$ are of
the same length
\newline
$ l((-1)^{(n-j)r}Q_{i_{1},i_{2},\cdots i_{k}}^{j}z^{-r}).$
\end{lemma}
{\bf Proof}

Lemma \ref{action} shows that if there is an edge going from $P$ to $Q$ then
$l(Q)=l(P)+1$. The lemma
now follows from the obvious remark that $l(1)=0$.

{\bf Q.E.D}

We are in a position to write down the integral (\ref{what_to_calc}).
Recall that \(W\) is the Weyl group of \(\widehat{\gtg}\) and
\(w=r_{m_{1}}r_{m_{2}}\cdots r_{m_{l}}\in W\) is a  decomposition (not
necessarily
reduced),where \(r_{m}\) denotes the  reflection at the corresponding simple
root
\( (\alpha_{m}) \).  \(m\) can be viewed as a map from
\(I_{1}(=\{1,2,\cdots,l\})\) to
\(I_{2}(=\{0,1,\cdots ,n\})\).Therefore,\(m^{-1}(j),\;j\in I_{2},\) is a subset
of
\(I_{1}\). Set,
\[\beta_{j}=\frac{2(r_{m_{l+2-j}}\cdots r_{m_{l}}\cdot
\lambda_{1},\alpha_{m_{l+1-j}})}
                  {(\alpha_{m_{l+1-j}},\alpha_{m_{l+1-j}})}+1,\;1\leq j\leq l
\]
and
\[ K_{w}(t_{1},t_{2},\cdots
,t_{l})=\prod_{j=1}^{l}\Gamma(-\beta_{j})^{-1}\times
\{\exp(-t_{l}F_{m_{l}}) \cdots
\exp(-t_{1}F_{m_{1}})\prod_{j=1}^{l}t_{j}^{-\beta_{j}-1}1\}, \]
where $1$ is viewed  as an element of $ V_{1}\otimes\cdots\otimes V_{N}$ equal
to the unit
function on the product of $N$ copies of the flag manifold.
With any path
\[\gamma:\;1 \stackrel{F_{j_{1}}}{\longrightarrow} Q_{1}
\stackrel{F_{j_{2}}}{\longrightarrow} Q_{2}
      \longrightarrow \cdots \longrightarrow Q_{r-1}
\stackrel{F_{j_{r}}}{\longrightarrow}
Q,\;r=l(Q)\] associate a polynomial in $t'$s:
\[f_{\gamma}(t)=\sum_{p_{1}<\ldots <p_{r},p_{i}\in
m^{-1}(j_{i})}t_{p_{1}}t_{p_{2}}\cdots
t_{p_{r}}.\]
( This is the only point where the decomposition \(w=r_{m_{1}}r_{m_{2}}\cdots
r_{m_{l}}\) enters the
calculation.)
Denote by $\Gamma^{j}$ the subgraph of $\Gamma$ consisting of all vertices
connected with $Q^{j}_{1}$
by an  oriented  path. It is equivalently defined as a subgraph generated by
all vertices
$(-1)^{(n-j)r}Q_{i_{1},i_{2},\cdots i_{k}}^{j}z^{-r}$ with the fixed
superscript $j$.
 Set
\begin{equation}
\label{form_kern_1}
 P_{w}^{j}(x,z;t_{1},t_{2},\cdots ,t_{l})=
   \sum_{l'=0}^{l}\;\sum_{Q\in\Gamma^{j}:l(Q)=l'}Q\sum_{\gamma\in
\cp(Q)}f_{\gamma}(t).
\end{equation}

\begin{theorem}
\label{th_form_kern_2}
\begin{equation}
\label{form_kern_2}
 K_{w}(t_{1},t_{2},\cdots t_{l})=\prod_{j=1}^{l}\Gamma(-\beta_{j})^{-1}
   \prod_{p=1}^{N}\prod_{j=1}^{n}\{P_{w}^{j}(x^{(p)},z_{p};t_{1},t_{2},\cdots
,t_{l})\}^{\mu^{(p)}_{j}}
   \prod_{j=1}^{l}t_{j}^{-\beta_{j}-1},
\end{equation}
where $\mu^{(p)}=(\mu^{(p)}_{1},\ldots,\mu^{(p)}_{n}),\;1\leq p\leq N$, is a
highest weight of
$V_{p}$ and $x^{(p)},\;1\leq p\leq N$, is a coordinate in the $p-$th copy of
the flag manifold.
 \end{theorem}

This theorem can be proved by induction on $l$ using Lemma\ref{action} and
Lemma\ref{length}.

 Let \({\cal M}\) be the local system of
continuous branches of \(K_{w}(t_{1},\cdots ,t_{l})\) over the Domain of
\(K_{w}(t_{1},\cdots ,t_{l})\)(say \({\cal D}\) ).Then finally we obtain

\begin{theorem}
\label{final_theorem_konec}
For any \(\sigma \in H_{l}({\cal M},{\cal D})\),the integral
\[\stackrel{\circ}{\Psi} \int_{\sigma}K_{w}(t_{1},t_{2},\cdots
,t_{l})dt_{1}dt_{2}\cdots dt_{l} \]
satisfies the system \( KZ(\lambda_{N+1},w\cdot \lambda_{1}) \)
\end{theorem}

{\bf Remark.} Theorem ~\ref{final_theorem_konec} gives solutions as an integral
over a certain
cycle depending on parameters $(x,z)$. These cycles belong to a homology group
of a complement
to a collection of hypersurfaces $K_{w}(t_{1},t_{2},\cdots ,t_{l})=0$ with
coefficients in a local
system defined over this complement. Note that generically ($l> 2$),
 and much unlike the case of Schechtman-Varchenko integral formulas,
$K_{w}(t_{1},t_{2},\cdots
,t_{l})=0$ is a union of hypersurfaces not isomorphic to hyperplanes and,
therefore, investigation of
the integral cannot be carried out by usual methods. We have already
encountered with the same
phenomenon in a different but related framework. As we argued in the
Introduction, our integral
formulas are intimately related to $\hgtg\mbox{ or }\gtg-$modules extended by
complex powers of a Lie
algebra generators. Rigorous treatment of such modules requires consideration
of a Lie algebra action
on sections of a local
system defined over a complement to -- highly non-linear -- set of ``shifted''
Schubert cells on a
flag manifold; for details see ~\cite{feig_mal}.

Note also that if $l=1,2$ then $K_{w}(t_{1},t_{2},\cdots ,t_{l})=0$ is
isomorphic to a union of
affine hyperplanes and the number of cycles can be calculated using results of
{}~\cite{sch_varch_2}.

  \subsection{{\bf Some examples}}

Theorem~\ref{th_form_kern_2} produces rather an algorithm to write down the
kernel of the integral
(~\ref{what_to_calc}) than a completely explicit formula for it:
(~\ref{form_kern_2})  relies on
(~\ref{form_kern_1}), while the latter is a linear combination of explicitly
given polynomials
$Q^{j}_{i_{1}\ldots i_{k}}z^{-r}$ with coefficients in the form
$\sum_{p}t_{p_{1}}\cdots t_{p_{r}}$
determined by the combinatorial data. We have been able to ``resolve'' the
combinatorial part of the
formula in the cases $\gtg=\gtsl_{2},\;\gtsl_{3}$. Although the
$\gtsl_{2}-$case was treated
in ~\cite{malff}, we discuss here both in a unified way for completeness.

{\bf The $\gtsl_{2}-$case.}
In this case the flag manifold is $\cP^{1}$, the big cell is
$\nc\subset\cP^{1}$. Fix a coordinate $x$
on $\nc$. Then the matrix $T$ ( via which the  polynomials $
Q^{j}_{i_{1},\ldots,i_{k}}$ are defined)
is given by $T=(x)$. The set of all   $ Q^{j}_{i_{1},\ldots,i_{k}}$ consists of
2 elements:
$Q^{1}=1,\;Q^{1}_{1}=x$. The vertices of the graph $\Gamma$ are all of the
form:
$A^{\epsilon}_{i}(x,z)=z^{-i}x^{\epsilon},\;\epsilon =0,1,\;i=0,1,2,3,\ldots$.
Further,
$\Gamma$ coincides with $\Gamma^{1}$ and is given by
\[1\stackrel{F_{1}}{\longrightarrow}x\stackrel{F_{0}}{\longrightarrow}z^{-1}
\stackrel{F_{1}}{\longrightarrow}z^{-1}x
\stackrel{F_{0}}{\longrightarrow}z^{-2}
\stackrel{F_{1}}{\longrightarrow} z^{-2}x\cdots.\]

 Observe that  the Weyl group of $\widehat{\gtsl}_{2}$ is a free group
generated by 2 reflections $r_{0},\; r_{1}$ and, therefore, each element is
uniquely expanded as
either
\[ r_{0}r_{1}\cdots\]
or
\[ r_{1}r_{0}\cdots\]
the second one being relevant to our calculation. Setting
\[w=\underbrace{r_{1}r_{0}\cdots r_{\epsilon}}_{l},\]

one obtains
\[P_{w}(x,z;t_{1},\ldots ,t_{l})\;(=P_{w}^{1}(x,z;t_{1},\ldots
,t_{l}))\;=\sum_{\epsilon =0}^{1}
\sum_{i=0}^{l-\epsilon}z^{-i}x^{\epsilon}\sigma_{i+\epsilon}(t_{1},\ldots
,t_{l}),\]
\[\sigma_{j}(t_{1},\ldots ,t_{l})=
\sum_{0\leq i_{1} <i_{2}<\cdots <i_{j}< l/2} t_{2i_{1}+1}t_{2i_{2}+1}\cdots
t_{2i_{j}+1},\]
completing the $\gtsl_{2}-$case.

{\bf The $\gtsl_{3}-$case.} The big cell is $\nc^{3}$ with coordinates $x_{1
1},x_{1 2},x_{2 2}$.
The matrix T is given by $T=\left(\begin{array}{ll}
                                       x_{1 1} & x_{1 2} \\
                                           1   & x_{2 2} \end{array} \right)$.
The set of all $Q_{i_{1},\cdots ,i_{k}}^{j}$ consists of 5 elements:
\[Q^{1}=1,\; Q_1^{j}=x_{j j}(j=1,2) \; Q_{1 1}^{1}=x_{1 1}x_{2 2}-x_{1
2},Q_2^2=x_{1 2} \].The
graph $\Gamma$ and its subgraphs $\Gamma^{1},\;\Gamma^{2}$ are given by
\[ \begin{array}{lllllllllllll}
\Gamma_{1}: & & &Q_{1}^{1}& \stackrel{F_{2}}{\longrightarrow}&Q_{1 1}^{1}&
            \stackrel{F_{0}}{\longrightarrow} &
-z^{-1}&\stackrel{F_{1}}{\longrightarrow}&
               -z^{-1}Q_{1}^{1}&\stackrel{F_{2}}{\longrightarrow}&-z^{-1}Q_{1
1}^{1}&\cdots \\
            & &\stackrel{F_{1}}{\nearrow}& & & & & & & & & &\\
            &1&                          & & & & & & & & & &\\
            & &\stackrel{F_{2}}{\searrow}& & & & & & & & & &\\
\Gamma_{2}: & & &Q_{1}^{2}&
\stackrel{F_{1}}{\longrightarrow}&Q_{2}^{2}&\stackrel{F_{0}}{\longrightarrow}
            & z^{-1}&\stackrel{F_{2}}{\longrightarrow}&z^{-1}Q_{1}^{2}&
             \stackrel{F_{1}}{\longrightarrow}&z^{-1}Q_{2}^{2}&\cdots \\
\end{array} \]
The Weyl group $W$ of $ \widehat{\gtsl_{3}} $ is realized as a group generated
by reflections at a
certain collection of affine lines on the plane ( see ~\cite{kac_book} ).
These lines produce a
covering of the plane by triangles, called {\em alcoves}, which $W$ acts on
effectively. Looking at
this action one obtains a collection of elements of $W$ so that a reduced
decomposition of any
element of $W$ is contained in it:

 Put $c:=r_{0}r_{1}r_{2}$($c$ is called a Coxeter element), then
any $w\in W$ can be written as
$w=sc^{k}tc^{-l}u$ where
$s=e,r_{2},r_{1}r_{2}$ , $u=e,r_{2},r_{2}r_{1}$
,$t=r_{0},r_{0}r_{1},r_{0}r_{1}r_{0}$ if $kl\neq 0$
and if $kl=0$, then $t$ can also be equal to $e$.

 Further one obtains
 \[ P^{j}_{w}(x,z;t_{1},t_{2},\cdots
,t_l)=\sum_{l'=0}^{l}A_{l'}^{j}f_{l'}^{j}(t),\;j=1,2, \]
 where
     \[ f_{l'}^{j}(t)=\sum_{p_1<p_2<\cdots <p_{l'},p_{i}\in m^{-1}((ji)_{3})}
                          t_{p_1}t_{p_2}\cdots t_{p_{l'}}, \]
$(k)_{3}\in\{0,1,2\}$ signifies the residue of $k$ modulo 3 and $m:\;\{1,\ldots
l\}\rightarrow
\{0,1,2\}$ is a function determining a reduced decomposition of $w$,
 \[A_{l'}^{1}=\left\{\begin{array}{ccc } (-z)^{-q}&\mbox{ if }&l'=3q\\
(-z)^{-q}x_{11}&\mbox{ if }&l'=3q+1\\
(-z)^{-q}(x_{11}x_{22}-x_{12})&\mbox{ if }&l'=3q+2,
\end{array}\right.\]

\[A_{l'}^{2}=\left\{\begin{array}{ccc } z^{-q}&\mbox{ if }&l'=3q\\
z^{-q}x_{22}&\mbox{ if }&l'=3q+1\\
z^{-q}x_{12}&\mbox{ if }&l'=3q+2.
\end{array}\right.\]


\begin{thebibliography}{99}

\bibitem{kn_zam}Knizhnik V., Zamolodchikov A., {\em Nucl.Phys.} {\bf B 247}
(1984) 83 - 103

\bibitem{ts_kan} A.Tsuchiya, Y.Kanie, {\em Adv. Stud. Pure Math.} {\bf 16}
297-372

\bibitem{sch_1}V.Schechtman, {\em International Mathematics Research Notices}
{\bf 3} (1992), 39-49


\bibitem{sch_varch_1}V.Schechtman V., A.Varchenko,  preprint,
Max-Planck-Institut fur Mathematik
MPI/89-51, 1989, {\em Letters in Math. Phys.} {\bf 20} 1990, 279-283


\bibitem{sch_varch_2}V.Schechtman , A.Varchenko, {\em Invent.Math.} {\bf 106},
139-194

\bibitem{kac_wak} V.G.Kac, M. Wakimoto ,{\em Proc. Nat'l Acad. Sci. USA} {\bf
85} (1988) 4956-4960


\bibitem{feig_sch_varch} B.Feigin, V.Schechtman, A.Varchenko,
{\em Letters in Math.Phys.} {\bf 20} (1990), 291-297

\bibitem{f_g_p_p_1} Furlan P., Ganchev A.Ch., Paunov R., Petkova V.B., {\em
Phys.Letters} {\bf 267}
(1991) 63-70

\bibitem{f_g_p_p_2} Furlan P., Ganchev A.Ch., Paunov R., Petkova V.B., preprint
CERN-TH.6289/91,
accepted for publication in {\em Nucl.Phys.B}


\bibitem{awata}Awata H.,Yamada Y.,  KEK-TH-316 KEK Preprint 91-209, January
1992

\bibitem{feig_mal} B.Feigin, F.Malikov, preprint RIMS-894 September 1992, to
appear in {\em Advances
in Sov.Math.}

\bibitem{feig_fren_0} B.Feigin ,  E.Frenkel ,
{\em Usp.Math.Nauk}
(={\em Russ.Math.Surv.}) {\bf 43} (1988)
227 - 228 ( in Russian )



\bibitem{bowkn} P.Bowknegt, J.McCarthy, K.Pilch,
 {\em
Progress of Theoretical Physics}, Supplement No.102 {\bf 70} (1988) 67-135

\bibitem{aw_ts_yam} H.Awata, A.Tsuchiya, Y.Yamada, {\em Nuclear Phys.} {\bf
B365} (1991) 680-696

\bibitem{malff}Malikov F.G., Feigin B.L., Fuchs D.B.,
{\em Funkc.Anal.i ego Pril.}(={\em Funct. Anal. Appl.}) {\bf 20}(1988) 2, 25-37
(in Russian)

\bibitem{kac_book}Kac V.G. {\em Infinite-dimensional Lie algebras},Cambridge
University Press 1990

 \end{thebibliography}
\end{document}